\documentclass[twocolumn,showpacs,preprintnumbers,amsmath,amssymb]{revtex4}
\usepackage{graphicx}% Include figure files
\usepackage{color}
\usepackage{dcolumn}% Align table columns on decimal point
\usepackage{bm}% bold math

\begin{document}

\title{Cantilever cooling with radio frequency circuits}

\author{D. J. Wineland, J. Britton, R. J. Epstein, D. Leibfried, R. B. Blakestad, K. Brown,
J. D. Jost, C. Langer, R. Ozeri, S. Seidelin, and J. Wesenberg
}

\affiliation{NIST, Time and Frequency Division,Boulder, CO}
\date{\today}

\begin{abstract}
We consider a method to reduce the kinetic energy in a low-order mode of a
miniature cantilever.  If the cantilever contributes to the
capacitance of a driven RF circuit, a force on the cantilever exists due
to the electric field energy stored in the capacitance.  If this
force acts with an appropriate phase shift relative to the motion of
the cantilever,
it can oppose the velocity of the
cantilever, leading to cooling.  Such cooling may enable reaching
the quantum regime of cantilever motion.
\end{abstract}
%\pacs{}

\maketitle

Precise control of quantum systems occupies the efforts of many
laboratories; an important recent application of such control is in
quantum information processing.  Some of this work is
devoted to controlling the motion of a mechanical oscillator at the
quantum level. This has already been accomplished in a ``bottom-up"
approach where a single atom is confined in a harmonic well.  For example,
it has been possible to make nonclassical mechanical oscillator
states such as squeezed, Fock, and Schr{\"o}dinger-cat states
\cite{meekhof96,monroe96_cats}.  However, for various applications,
there is also interest in a ``top-down" strategy, which has
approached the quantum limit by using smaller and smaller
micro-mechanical resonators (for a summary, see e.g.,
\cite{schwab05}).  In this case, small ($\sim 1~\mu$m) mechanical resonators, having
low-order mode frequencies of approximately 10 - 100 MHz, can approach the quantum regime
at low temperature ($<$ 1 K); mean thermal
occupation numbers of approximately 50 have been achieved
\cite{lahaye04}.

To reach the quantum level of a mechanical oscillator, an efficient
cooling mechanism is desirable.  With harmonically bound atoms,
this can be achieved with laser cooling where, in a room temperature
apparatus, the modes of mechanical motion can be cooled to a level
where the occupation numbers $\langle n \rangle$ of the quantized
modes reach values less than 0.1 for oscillation frequencies $\sim 1
- 10$ MHz~\cite{diedrich89,monroe_cooling}.  For more macroscopic
mechanical oscillators, other means are sought.  An
extension of laser cooling of atoms would be to couple laser cooled
atomic ions to a (charged) macroscopic oscillator
\cite{heinzen90,bible,tian04,hensinger05}; here, the macroscopic
oscillator would be cooled sympathetically through its Coulomb
coupling to the ions.  Analogously, the resonator might be cooled by
coupling to other quantum systems
\cite{wilson-rae04,martin04,zhang05,blencowe05,clerk05}.

Cooling of a macroscopic mechanical oscillator can also be achieved by
feedback applied with optical forces.  The feedback can be
obtained using external electronics to control radiation pressure as
in \cite{cohadon99}.  In \cite{metzger04}, the authors describe
several passive means of optical feedback.  Their experiment reported
cooling by means of photothermal forces, but they also describe
theoretically passive-feedback cooling by means of the radiation
pressure force.  In this note we describe (classically) a related
possible cooling mechanism where the cooling force is between
capacitor plates that contain a radio frequency (RF) electric field.
\begin{figure}%[t]
\includegraphics[width=8cm]{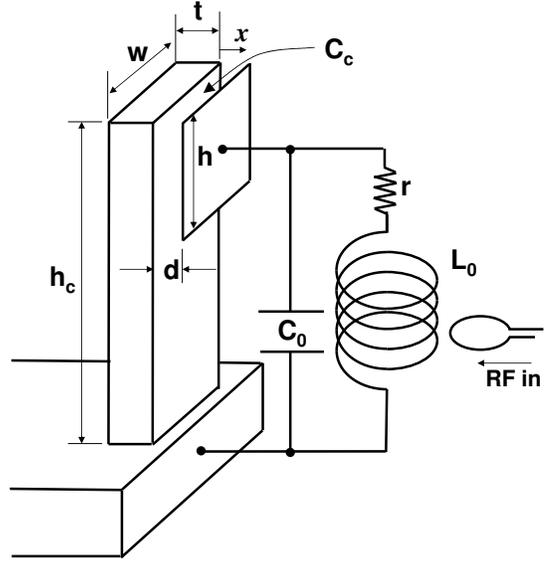}
\caption{Cantilever and associated RF circuitry. We assume the
cantilever is approximated by a thin beam fixed to a rigid base at
one end (lower left).  The plate of area w$\cdot$h attached to the RF circuit is also assumed to
be mechanically rigid.}\label{cantilever_circuit}
\end{figure}

To describe the cooling mechanism, we refer to the simplified situation
shown in Fig.~\ref{cantilever_circuit}.  We assume a conducting beam
cantilever, having density $\rho$, which is fixed rigidly at one end. One
face of the cantilever is placed a distance d from a rigidly mounted plate of
area w$\cdot$h thereby forming a parallel-plate capacitor $\mathrm{C_c} = \epsilon_0$w$\cdot$h/d
where $\epsilon_0$ is the vacuum dielectric constant.  An inductor
L$_0$ and capacitor C$_0$ are connected in parallel with
$\mathrm{C_c}$ to form a parallel tank circuit with (RF) resonant
frequency $\Omega_0 = 1/\sqrt{\mathrm{L_0(C_0 + C_c)}}$.
For simplicity, we assume all losses in the RF
circuit (including the coupling to the source impedance) are
represented by a resistance r.  We also assume
 $Q_{\mathrm{RF}} = \Omega_0 \mathrm{L_0/r} \gg 1$ where
$Q_{\mathrm{RF}}$ is the quality factor of the RF circuit.

For simplicity, we consider only the lowest-order bending mode of the
cantilever, where the free end oscillates back and forth (in the $\hat{x}$ direction
in Fig.~\ref{cantilever_circuit}) with frequency $\omega_c$, which we assume to be much
smaller than $\Omega_0$. Small displacements $x$ of the
cantilever can be described by the equation of motion
\begin{equation}
m\ddot{x} + m \Gamma \dot{x} + m \omega_c^2 x = F,
\label{equation_of_motion}
\end{equation}
where $\Gamma$ is the damping rate of the cantilever oscillation,
$F$ is the force on the end of the cantilever, and $m$ is the
effective mass of the cantilever, given by $\rho \mathrm{w h_0 t/4}$~\cite{sidles95}.
The force $F$ includes random thermal forces as well as
purposely applied forces.

If a potential V is applied to capacitor $\mathrm{C_c}$, the
capacitor plates experience a mutual attractive force; in the
context of Fig.~\ref{cantilever_circuit}, the cantilever feels a
force $F_{\mathrm{E}} = \epsilon_0 \mathrm{E^2 w h}/2 = \mathrm{C_c
V^2/(2 d)}$ in the positive $x$ direction, where E = V/d is the
electric field between the capacitor plates.  Here we will be
interested in the case where V is an applied RF potential
$\mathrm{V_{RF} \cos \Omega_{\mathrm{RF}} t}$ with $\Omega_{\mathrm{RF}} \simeq \Omega_0$. Since $\omega_c \ll
\Omega_0$, the force for frequencies around $\omega_c$
is given by the averaged RF force

\begin{equation}
F_{\mathrm{E}} = \frac{\mathrm{C_c} \langle \mathrm{V^2} \rangle}{2 \mathrm{d}} =
\mathrm{\frac{C_c V_{RF}^2}{4 d}} = \frac{\epsilon_0 \mathrm{w h V_{RF}^2}}{\mathrm{4
d^2}}.\label{capacitor_force}
\end{equation}

This RF capacitive force will give rise to the cooling as follows.
As the cantilever oscillates back and forth, its motion modulates
the overall capacitance of the RF circuit ($\mathrm{C_c + C_0}$)
 thereby modulating the RF circuit's resonant
frequency.  If the input RF frequency is tuned to the lower side of the RF
resonance, as the frequency of the circuit is modulated, so too is
the RF electric field amplitude in the capacitance of the circuit.  As described below, this gives
rise to an additional oscillating capacitive force term that will shift the
resonant frequency of the cantilever.  However, due to the finite
response time of the RF circuit (given by $Q_{\mathrm{RF}}$), there is a
phase lag in this additional capacitive force term relative to the cantilever motion.
This phase lag leads to a force component that opposes the velocity
of the cantilever, thereby leading to the cooling.

The average RF capacitive force will displace the equilibrium position of
the cantilever. However, we will be mainly interested in small
deviations of the cantilever around its equilibrium position
$\mathrm{d_0}$; therefore, for the moment, we will assume this
displacement is absorbed into the definition of $\mathrm{d_0}$ and
write d $\equiv \mathrm{d_0} - x$~\footnote{This expression for d
neglects the curvature of bending mode and is strictly true only for
h $\ll \mathrm{h_c}$ and $\mathrm{d_0 \ll w, h}$.}. For small deviations $x$ around the
equilibrium position, we write
\begin{equation}
\delta F_{\mathrm{E}} = (\partial F_{\mathrm{E}}/\partial x)
x.\label{force}
\end{equation}
To evaluate this expression, we note that
$F_{\mathrm{E}}$ depends on $x$ through d as well as
through $\mathrm{V_{RF}}$,
\begin{equation}
\frac{\partial F_{\mathrm{E}}}{\partial x} = \frac{\epsilon_0 \mathrm{w
h}}{4} \biggl[\mathrm{V_{RF}^2} \frac{\partial
\mathrm{(1 / d^2)}}{\partial x} + \frac{1}{\mathrm{d^2}} \frac{\partial
\mathrm{V_{RF}^2}}{\partial x} \biggr].\label{force2}
\end{equation}
This is because as $x$ changes, $\mathrm{C_c}$ changes thereby
changing $\Omega_0$. If RF is applied to the circuit of
Fig.~\ref{cantilever_circuit} at a frequency $\Omega$ near
$\Omega_0$, $\mathrm{V_{RF}}$ will depend on $x$ due to its
dependence on $\Omega - \Omega_0$.

Here, we will assume that the RF frequency modulation caused by the
cantilever is less than than the bandwidth
$\Omega_0/Q_{\mathrm{RF}}$ of the RF circuit. In this case we can write
\begin{equation}
\frac{\partial \mathrm{V_{RF}^2}}{\partial x} \simeq \frac{\partial
\mathrm{V_{RF}^2}}{\partial \Omega_0} \cdot \frac{\partial
\Omega_0}{\partial \mathrm{C_c}} \cdot \frac{\partial
\mathrm{C_c}}{\partial x}\ .\label{pwr_vs_x}
\end{equation}
The first factor on the right-hand-side of this equation can be
obtained from the expression for the RF potential across the circuit
of Fig.~\ref{cantilever_circuit} relative to the maximum
(on-resonance) RF potential $\mathrm{V_{max}}$ for a fixed input
power
\begin{equation}
\mathrm{\frac{V_{RF}^2}{V_{max}^2}} \simeq \frac{1}{1 + \bigl[2
Q_{\mathrm{RF}} \frac{\Omega_{\mathrm{RF}} - \Omega_0}{\Omega_0}
\bigr]^2}\ .
\end{equation}
When the input
RF frequency is tuned to the ``half-power" points of the RF circuit ($2
Q_{\mathrm{RF}} (\Omega_{\mathrm{RF}} - \Omega_0)/\Omega_0 = \pm
1$), we find $\partial \mathrm{(V_{RF}^2 / V_{max}^2)} /\partial
\Omega_0 = \pm Q/\Omega_0$.

The second and third factors
in Eq.~(\ref{pwr_vs_x}) are $\partial \Omega_0/\partial
\mathrm{C_c} = -\Omega_0/(2 \mathrm{(C_0 + C_c)})$ and $\partial \mathrm{C_c} / \partial x =
\mathrm{C_c / d_0}$. With these expressions, Eq.~(\ref{pwr_vs_x}) at the half-power points becomes
\begin{equation}
\frac{\partial \mathrm{V_{RF}^2}}{\partial x} = \pm
\mathrm{V_{max}^2} \frac{Q_{\mathrm{RF}}}{2 \mathrm{d_0}}
\frac{\mathrm{C_c}}{\mathrm{C_c + C_0}}\label{pwr_vs_x_2},
\end{equation}
where the $+~(-)$ sign refers to $\Omega_{\mathrm{RF}} <
 (>)\ \Omega_0$.  This dependence of the capacitor plate force with
 cantilever position is analogous to the dependence of the radiation pressure
 force on the mirrors in an optical cavity with the spacing of the cavity mirrors~\cite{metzger04}.

 Combining these expressions, Eq.~(\ref{force2}) becomes
 \begin{equation}
 \delta F_{\mathrm{E}} = \mathrm{\frac{C_c V_{max}^2}{4 d_0^2}} \biggl[1 \pm \frac{Q_{\mathrm{RF}}}{2}
 \mathrm{\frac{C_c}{C_c + C_0}}
 \biggr] x. \label{force_2}
 \end{equation}
 When substituted into the right-hand side of Eq. (\ref{equation_of_motion}), this
 expression only alters the spring constant $m \omega_c^2$ and
 therefore the oscillation frequency of the cantilever.  However, Eq.
 (\ref{pwr_vs_x_2}) gives the change of
 $\mathrm{V_{RF}^2}$ vs. $x$ assuming that the RF energy has reached its steady-state value.
 In fact, as $x$
 changes, $\mathrm{V_{RF}^2}$ requires a time $\tau$ to reach its
 steady state value, where $\tau$ is the decay time of the RF
 circuit.

 This lag in time is analogous to how the output voltage
 $\mathrm{V_{out}}$
 following a low-pass RC filter changes in response to changes in
 the input voltage  $\mathrm{V_{in}}$.  For this case, the output voltage responds to
 input signals $\mathrm{V_{in} = |V_{in}}|e^{i \omega t}$ with a
 transfer function
\begin{equation}
\mathrm{V_{out}/V_{in}} = \cos{\phi}\ e^{-i \phi},
\end{equation}
where $\phi \equiv \tan^{-1}(\omega \tau)$ and
$\tau =$ RC.

Analogously, a similar transfer function must be
applied to Eq.~(\ref{pwr_vs_x_2}) and the second term in Eq.
(\ref{force_2}),
assuming $x \rightarrow x_0 e^{i \omega t}$.
With this
modification, Eq.
(\ref{force_2}) becomes
 \begin{equation}
 \delta F_{\mathrm{E}} = \mathrm{\frac{C_c V_{max}^2}{4 d_0^2}} \biggl[1 \pm \cos{\phi} e^{-i \phi} \frac{Q_{\mathrm{RF}}}{2}
 \mathrm{\frac{C_c}{C_c + C_0}}
 \biggr] x, \label{force_3}
 \end{equation}
where $\omega$ is the cantilever frequency and $\tau$ is now the decay time of the RF circuit.
Noting that $i x = \dot{x}/\omega$, Eq.~(\ref{equation_of_motion}) with the force
$F = \delta F_{\mathrm{E}} + F_0 e^{i \omega t}$ becomes
\begin{equation}
m\ddot{x} + m (\Gamma + \Gamma')  \dot{x} + m \omega_c^2 (1 -\kappa)
x = F_0 e^{i \omega t}, \label{equation_of_motion_2}
\end{equation}
where
\begin{equation}
\kappa \equiv \frac{\mathrm{C_c V_{max}^2}}{4 m \omega_c^2
\mathrm{d_0^2}} \biggl[ 1 \pm \frac{\cos^2{\phi}\ Q_{\mathrm{RF}}
\mathrm{C_c}}{\mathrm{2 (C_c + C_0)}} \biggr],
\end{equation}
\begin{equation}
\Gamma' \equiv \pm \frac{Q_{\mathrm{RF}}
\mathrm{V_{max}^2 C_c^2}}{16 m \omega_c \mathrm{d_0^2 (C_c + C_0)}} \sin{2 \phi},
\end{equation}
and the $\pm$ sign conventions are as noted above. The term $\kappa$
results in a frequency shift of the cantilever mode.  For
$\Omega_{\mathrm{RF}} < \Omega_0$ the term $\Gamma'$ gives rise to
an increased damping and will lead to cooling.

Before considering cooling, we must first examine the sources of
noise in the system. Assuming the cantilever is at temperature $T$,
the spectral density of force fluctuations acting on the
isolated cantilever at frequencies near $\omega_c$ is given by~\cite{sidles95}.
\begin{equation}
S_F(\mathrm{CANT}) = 4 k_B T m \omega_c/Q_c = 4 k_B T m/\tau_c,
\end{equation}
where $k_B$ is Boltzmann's constant, and $Q_c$ and $\tau_c$ are the
cantilever $Q$-factor and (energy) decay time constant.
%\footnote{This expression can be checked by integrating the
%cantilever response to this force (Eq.~(\ref{equation_of_motion}))
%over all frequencies to obtain $m \langle v^2 \rangle /2 = k_B
%T/2$}.\mathrm{C_c}
%

We must also consider noise in the RF circuit and its effect on the
cantilever. In Eq.~(\ref{capacitor_force}), we need to replace
$\mathrm{V_{RF}}$ with $\mathrm{V_{RF}}+ v_n(\mathrm{C_c})$, where $v_n(\mathrm{C_c})$ is the
Johnson noise potential (characterized by noise spectral
density $S_{v_n}(\mathrm{C_c})$)) across the cantilever capacitance
$\mathrm{C_c}$ due to resistance in the RF circuit. In particular,
the cantilever will be affected by RF noise at frequencies near
$\Omega_{\mathrm{RF}}
\pm \omega_c$ because cross terms in Eq.~(\ref{capacitor_force})
will give rise to random forces at the cantilever frequency.
(Here we assume the RF modulation index due to cantilever motion is
much less than one, which will be true for the examples below.)
The voltage noise spectral density from the resistor $r$ in Fig.~\ref{cantilever_circuit}
 is given by $S_{v_n}(r) =  4 k_B T r$.  From an analysis of
 the circuit we can then calculate $S_{v_n}(\mathrm{C_c})$.
With this and
Eq.~(\ref{capacitor_force}), we find the spectral density of force
fluctuations due the RF circuit noise to be
\begin{equation}
S_F(RF) = \frac{1}{2} \biggl[ \mathrm{\frac{C_c V_{RF}}{d_0}}
\biggr]^2 S_{v_n}(\mathrm{C_c}).
\end{equation}
In general, we must also consider other sources of noise, such as that from a
detection circuit that is connected to the circuit in
Fig.~\ref{cantilever_circuit}.  Depending on the detection method
used, this noise can be important; however, for simplicity, we assume the
detection can be switched on and off without significantly affecting the energy of
the cantilever.

Considering only $S_F(\mathrm{CANT})$ and $S_F(\mathrm{RF})$, the
effective temperature of the mode that is acted on is increased
by the additional noise from the RF circuit, but lowered by the
increased damping
\begin{equation}
\frac{T_{\mathrm{eff}}}{T} = \biggl[ \frac{\Gamma}{\Gamma + \Gamma'}\biggr] \biggl[
\frac{S_F(\mathrm{RF}) + S_F(\mathrm{CANT})}{S_F(\mathrm{CANT})}
\biggr].
\end{equation}

To get an approximate idea of the cooling that might be achieved, we
first consider a silicon cantilever where the doping is high enough
that we can neglect heating from RF currents. We assume
$\mathrm{h_c} = 1.5$ mm, $\mathrm{h} = 0.5$ mm, $\mathrm{t} = 20\
\mu$m, $\mathrm{w} = 400\ \mu$m, $\mathrm{d_0} = 10\ \mu$m.  The
frequency of the lowest-order bending mode is given
by~\cite{sidles95}
\begin{equation}
\omega_\mathrm{c} = 3.516 \mathrm{\frac{t}{h_\mathrm{c}^2}}
\sqrt{\frac{E}{12 \rho}},
\end{equation}
where $E = 1.07 \times 10^{11}$ Pa and $\rho = 2.33 \times 10^3\
\mathrm{kg/m}^2$ are the Young's modulus and density of silicon.  For these parameters we find
$\omega_\mathrm{c}/2 \pi = 9.73$ kHz, $m = 7.00 \times 10^{-9}$ kg, and
$\mathrm{C_c} = 0.177$ pF. We assume $\tau_\mathrm{c} = 5$ s.

For the RF circuit we assume $\Omega_{\mathrm{RF}}/2 \pi = 50$ MHz,
$\mathrm{C_0} = 10$ pF, $Q_{\mathrm{RF}} = 400$. With
$\mathrm{V_{max}} = 20$ V, we find
$S_F(\mathrm{RF})/S_F(\mathrm{CANT}) = 1.40$, $\kappa
= 0.0302$, $\Gamma '/\Gamma = 558$, and $T_{\mathrm{eff}}/T = 4.30 \times 10^{-3}$.
For this example, the final
mean occupation number $\langle n \rangle$ of the fundamental mode of cantilever is $>
10^6$ for T = 300 K.
Assuming the cantilever
spring constant is given by $m \omega_\mathrm{c}^2$~\cite{sidles95},
the deflection $\Delta x$ of the cantilever due to the mean RF force
is obtained from $m \omega_\mathrm{c}^2 \Delta x = \mathrm{C_c
V_{RF}^2/4 d_0}$.  For the parameters here, we find $\Delta x / \mathrm{d_0} =
3.39 \times 10^{-3}$.

We can also consider coupling a cantilever to a high-Q
stripline resonator.  For simplicity, we choose a 1/4-wave resonator where the high impedance end of
the stripline and the end of the cantilever form
 a capacitor of area $\mathrm{w \cdot h}$ and plate
spacing $\mathrm{d_0}$ similar to the case in Fig. \ref{cantilever_circuit}.
The equivalent capacitance $\mathrm{C_0}$ becomes $\pi/(4 \omega_0 Z_0)$ where
$Z_0$ is the characteristic impedance of the
the line.  Assuming again the characteristics of a doped silicon resonator with $\mathrm{h_c} = 9\ \mu$m, $\mathrm{h} = 3\ \mu$m,
 $\mathrm{t} = 1\
\mu$m, $\mathrm{w} = 10\ \mu$m, $\mathrm{d_0} = 0.1\ \mu$m, $\tau_\mathrm{c} = 0.4$ ms,
and RF parameters $\Omega_{\mathrm{RF}}/2 \pi = 25$ GHz,
$\mathrm{C_0} = 0.1$ pF, (characteristic impedance $Z_0 = 50\ \Omega$), $Q_{\mathrm{RF}} = 500$,
and $\mathrm{V_{max}} = 10$ V, we find $\omega_\mathrm{c}/2 \pi = 13.5$ MHz,
$m = 5.24 \times 10^{-14}$ kg, $\mathrm{C_c} = 0.00266$ pF,
$S_f(\mathrm{RF})/S_f(\mathrm{CANT}) = 2.09$, $\kappa
= 0.123$, $\Gamma '/\Gamma = 972$, $T_{\mathrm{eff}}/T = 3.18 \times 10^{-3}$, and $\Delta x / \mathrm{d_0} =
8.78 \times 10^{-3}$.  If we assume $T = 50$ mK, this would imply a
mean occupation number of the cantilever $\langle n \rangle < 1$, necessitating a fully
quantum treatment~\footnote{S. Girvin, private
communication.}.

Of course, variations on this basic layout should be considered.
Different materials need to explored and for mechanical robustness,
it might be better to fix the cantilever at both ends.  Multiple
modes could be cooled with the same configuration provided that the
cantilever motion provides sufficient modulation of the RF circuit
frequency. For the quantum limit of cooling, the important case
where $\omega_c \gg \Omega_0/Q_{\mathrm{RF}}$ must also be
considered~[20]. The unstable regime $\Gamma ' = - \Gamma$ where the
cantilever breaks into self-oscillation would also be interesting
and could provide a further check of the model parameters.

We thank S. Girvin, J. Moreland, and S.-W. Nam for helpful comments.

\end{document}